\documentclass[11pt,a4paper]{article}
\usepackage{graphicx}
\usepackage[left=2cm,right=2cm,top=2cm,bottom=2.5cm]{geometry}

\newcommand {\vJ}{\vec{J}}
\newcommand {\vh}{\vec{h}}

\newcommand {\cG}  {{\cal G}}
\newcommand {\cD}  {{\cal D}}
\newcommand {\cP}  {{\cal P}}
\newcommand {\cF}  {{\cal F}}
\newcommand {\cT}  {{\cal T}}
\newcommand {\cH}  {{\cal H}}
\newcommand{\dd}{\mathrm{d}}
\newcommand{\con}{\setlength{\itemsep}{0cm}\setlength{\parskip}{0cm}}
\begin{document}
\title{The marginally stable Bethe lattice spin glass revisited}
\author{Giorgio Parisi\\Dipartimento di Fisica,
 Universit\`a di Roma {\it La Sapienza},
\\
INFN, Sezione di Roma I,  CNR-NANOTEC UOS Roma \\
P.le A. Moro 2, I-00185 Roma, Italy
}
\date{}
\maketitle

\abstract{
Bethe lattice spins glasses are supposed to be marginally stable, i.e. their equilibrium probability distribution changes discontinuously when we add an external perturbation. So far the problem of a spin glass on a Bethe lattice has been studied only using an approximation where marginal stability is not present, which is wrong in the spin glass phase. Because of some technical difficulties, attempts at deriving a marginally stable  solution have been confined to some perturbative regimes, high connectivity lattices or temperature close to the critical temperature. Using the cavity method, we propose a general non-perturbative approach to the Bethe lattice spin glass problem using approximations that should be hopefully consistent with marginal stability\footnote
{I would like to dedicate this paper to the memory of my good friend Leo Kadanoff. I  presented for the first time my spin glass theory at a Les Houches winter workshop in 1980 in an after dinner seminar. Leo was among the public,  quite near to me in the front row, and I was comforted by his facial expressions of interest and approbation during the seminar; Leo congratulated with me after the seminar. His reactions were an important confirmation that I was on the right track: I was very grateful to him. }.}

\section{Introduction} 
The existence of multiple equilibrium states is the distinctive hallmark of glassiness \cite{MPV,parisibook2}. While at finite temperature these equilibrium states may be metastable (or maybe not, depending on the system), the issue of metastability becomes irrelevant in the zero temperature limit.  A crucial question is how these states differ one from the other and how they are distributed in the phase space (the phase space becomes an infinite dimensional space in the thermodynamic limit). Moreover, in this situation a small perturbation produces a large rearrangement of the relative free energy of the equilibrium states: after a perturbation, the lowest equilibrium state may be quite different from the previous one. The linear response theorem has to be modified in order to take care of the existence of many equilibrium states.

There are two main possible scenarios that have quite different properties. 

\begin{itemize} \con
\item Different equilibrium states are  scattered in phase space in a nearly randomly fashion: they stay at a minimum distance one from the other. In this situation, if you live inside one equilibrium state, you do not feel the existence of other equilibrium states: the barriers are very high \cite{MPV}. We will call this scenario the {\sl stable glass}.
\item The distribution of equilibrium states in phase space is very far from a random one. Each state is surrounded by a large number of other equilibrium states that are arbitrarily near:  the equilibrium states form a fractal set. The barriers for going from one state to another may be very small. 

If you live in one equilibrium state, you feel that  you can move in many directions (typically in the direction of nearby states) without increasing too much the free energy: in other words, there are nearly flat directions in the potential  exactly like at a second order transition critical point \cite{MPV}.  The spectrum of small oscillations within an equilibrium state has an excess of low-frequency modes that in some case may dominate the one coming from other more conventional sources (e.g. phonons or magnons). The precise form of this anomalous low-frequency spectrum and the localization properties of the eigenvalues crucially depend on the details of the theory. These glassy systems are self-organized critical systems.  We call this scenario the {\sl marginally stable glass}.  
\end{itemize}

There are differences among the two scenarios that are quite important:
\begin{itemize} \con
\item Let us consider the correlation functions of local quantities that are not conserved. In the stable phase, they decay exponentially in both space  and time. In the marginal stable phase, the  correlations decay as a power both in space and in time.
\item Nonlinear susceptibilities are finite in the stable glass phase and they are often infinite in the marginal stable phase.
\item In both cases, the application of a finite external perturbation may force the system to jump from one equilibrium state to another quite different state.  This behavior is usually called chaotic because the set of equilibrium states has a chaotic dependence on the parameters of the Hamiltonian.

The two scenarios are different if we consider what happens in the time domain when we apply the perturbation at a given time. It can be argued that the time to reach equilibrium is exponentially large for the stable glasses because the decay must happen via a thermally activated process. On the contrary, this equilibration time should be not so large in the marginally stable scenario where the barriers are much smaller.
\end{itemize}

In the last forty years, a mean field theory of glassy systems has been constructed   using the replica approach \cite{MPV}, that I will not describe here. Most of the results of the replica approach have been reproduced using a more intuitive probabilistic approach that, at the time being,  can be used as a starting point for a rigorous analysis. Occasionally I will  use the replica terminology that I will  recall in the following lines.

I order to discuss the possible scenarios it is convenient to introduce a distance $d$ and an overlap $q$ among equilibrium states. Usually, they are defined in such a way that $q\equiv 1-d$, the distance $d$ being in the range $d_m\le d\le d_M$: two states are at distance  $d_m$ only if they coincide.

\begin{itemize} \con
\item In the high-temperature phase, where there is only on equilibrium state,  the replica symmetry is exact. In this case, $d_m=d_M$ and the role of the distance is quite  trivial.
\item When multiple equilibrium states do exist, replica symmetry is spontaneously broken (RSB). It is possible to construct a finer classification.
\begin{itemize} \con
\item The overlaps among states can take only two values (i.e. one step RSB): we stay in stable glass phase. 
\item The overlaps among states can take only $r+1$ values ($r$ steps RSB): we should also stay in stable glass phase. As far I know this situation is very rare.
\item The overlaps among states can take all possible values in the interval $[q_m:q_M]$ (continuous RSB): we stay in the marginal glass phase.
\end{itemize}
\end{itemize}
 
In the previous discussion, we have argued that  the continuous RSB scenario implies marginal stability. However, there are subtle differences between RSB and marginal stability.
\begin{itemize}
\item In the continuous RSB scenario \cite{MPV} different states give a similar contribution to the partition function, apart from multiplicative factors of order one, i.e. there are states whose {\sl extensive} free energies differ of quantities that are of order one.
\item The marginal stability scenario is slightly more general. There is no requirement that the  {\sl extensive} free energies differences should be of order one: they could diverge with the volume, however, the difference in the intensive free energies should always become zero in the thermodynamic limit. 
\end{itemize}

The concept of marginally stable glasses is quite old\footnote{
The expression {\sl marginal stability} was also used to characterize some off-equilibrium metastable states \cite{KWT} with a slightly different meaning.}. More than thirty years ago \cite{MPV} it was argued that in the infinite range model for spin glasses (i.e., the Sherrington-Kirkpatrick model \cite{SK}, SK in short) all the aforementioned properties of marginally stable glasses are present \cite{DK0}. This model can be exactly solved using the appropriate mean field theory \cite{GuerraI,Tala,TalaBook,PanBook,ContucciLibro}.

For finite dimensional spin glasses, the situation is less clear, also because we have a limited command of the perturbation theory around mean-field theory.  When the  magnetic field is zero, numerical experiments in dimensions 3 or higher are in very good and detailed agreement with theoretical predictions of mean field theory. For example in finite dimensions marginal stability predicts the existence of long-range correlations \cite{DK1} that are clearly observed \cite{Janus}.  

Ten years ago it was suggested that also some two and three-dimensional structural glasses, in particular, hard sphere systems at infinite pressure (i.e., in the jamming limit), are marginally stable, in the restricted sense that they are unstable under an infinitesimal perturbation \cite{WYART}. Quite recently \cite{LARGED1,FINAL} a mean field model of hard spheres has been constructed and solved in the infinite dimensional limit ($d\to\infty$, where $d$ is  the dimension  of the space where the spheres  move). The model has many features similar to the SK model:  all the stigmata of marginal stability are present here, suggesting that a similar situation also holds for some finite-dimensional glasses.

These advances have renewed the interest on the analytic study of marginally stable systems. Unfortunately manageable (I do not dare say simple) formulae exist only in the case where the effective number of neighbors of a given degree of freedom goes to infinity. In this case, the number of forces acting on a given degree of freedom becomes infinite: the central limit theorem can be applied and some crucial quantities become Gaussian. When each degree of freedom interacts only with a finite number of other degrees of freedom, the distributions are not Gaussian: also in the mean field case, we lack explicit manageable formulae.

In this paper, we concentrate our attention on spin glasses  on Bethe lattice both for finite coordination and in the infinite coordination limit, the SK model, where simple mean-field theory  is correct (more or less by definition). These two cases are among the most studied mean field models. However, our understanding of Bethe spin glasses is far from being satisfactory. A simple way to compute the free energy consists in using the formulae written in \cite{MP2} for the one step RSB, that can be easily generalized to $r$ steps RSB: finally we send $r$ to infinity. As we shall see later, this computation is quite complex for finite $r$; moreover we have no proof that it gives the correct result in the limit $r\to\infty$, i.e. in the continuous RSB.

Generally speaking, we expect we can reasonably approximate the free energy of a Bethe lattice spin glass with the one step computation (i.e. $r=1$). However, this is not always true. For example in the low-temperature limit in the SK model, the specific heat is proportional to $T^2$ (i.e. the correct behavior) and it is proportional to $T$ if we have a finite number of RSB steps. 

However, the most dramatic differences among the approximate treatment of RSB (e.g. finite number of steps)  and the exact solution (quite likely with continuous RSB) come from those quantities that have very different properties in the marginally stable phase, as the spectrum of small oscillations, nonlinear susceptibilities and so on. 

In order to proper study these quantities we need to have under good control the formulae for the free energy for continuous RSB: indeed many of the properties that characterized  marginal stability are related in a direct way to the continuous breaking of replica symmetry. Using the replica jargon\footnote{I apologize for using here the replica jargon; this could be avoided in many cases because the replica language statements have been translated into the probabilistic language in most of the cases.  Unfortunately, the stability analysis based on infinitesimal perturbations is not yet been translated into a probabilistic language and I cannot avoid using the replica language.}  these properties are the effect of the breaking the invariance of the replica Hamiltonian under infinitesimal permutations, a phenomenon that appears only in the case of continuous RSB.

The aim of this paper is to review the present  understanding of  marginally stable systems in mean field theory and to propose a possible path forward. The review part of the paper is relatively long, but it  is needed: many progresses have been done in the recent years  \cite{Ruelle,Bolt,GuerraI,FL,ASS,PANU,PanBethe}, but many of the most recent beautiful results are not widely known by  physicists. 

The marginally stable scenario has proved to be correct for spin glass, but  similar proofs are available for many other models: exact results can be obtained for the Potts model \cite{Potts}, for the vector spin models \cite{Vector}, for the  $K$-sat model with large number of clauses \cite{PANK}, for the multi-species Sherrington-Kirkpatrick model  \cite{MS,MS1} and for the temperature chaos \cite{Chaos,Chaos1}.

The plan of the paper is the following: in the second section we define Bethe lattice spin glasses, in the next section we show how to write the mean field (cavity) equations in presence of multiple states, finally, in the last section, we present our proposal that is constructed by generalizing the known solution of the SK model. 
\section{Ising spin glasses  on a Bethe lattice}

In the thirties,  Bethe invented the Bethe (cavity) approximation for the ferromagnetic Ising spin models. He assumed that when one removes a spin in a point $i$, forming a cavity in $i$,  the nearby spins become uncorrelated, apart from a possible global magnetization. Of course this approximation in not exact in finite dimensional spin glasses: it is exact only in one dimension and in the infinite dimensional limit: in this case, it reduces to the standard mean-field approximation. 

The approximation is not exact (in dimensions greater than 1) because,  when the cavity is formed, the spins at the border of the cavity (who do not have a direct interaction, at least in a square or in a simple cubic lattice)  are correlated due to the existence of closed paths on the lattice.

A graph $\cG$ is called a {\it Bethe lattice} if  in the case of  an Ising model with nearest neighbor interaction the  Bethe approximation becomes exact in the thermodynamic limit, i.e., when  the number $N$ of points goes to infinity (see for example \cite{MP}). Usually, a sufficient requisite for the exactness  of the  Bethe approximation is that in this limit the graph becomes locally a tree: the number of finite length loops should be negligible.  More precisely: the probability that a node belongs to a loop of length equal or less that $d$ should go to zero when $N \to \infty$ at fixed $d$.

In this paper we will consider one of the most studied cases: the graph $\cG$ is a random representative of the Erd\"os-R\'enyi  ensemble $G(N,M)$. This ensemble contains all possible graphs with $N$ nodes and $M$ edges: this is the so-called the Viana Bray model \cite{VB}\footnote{ Similar studies can be done on random regular lattices, i.e. on the fixed connectivity lattice studied in many papers, e.g.  \cite{FA,MP0,DDM}.}. The $N \to \infty$ limit  is taken keeping $M= N z/2$:  the average number of nearest neighbors of a given node is  $z$.

Let me briefly recall some of the main features of the typical graph of this ensemble for large $N$.
\begin{itemize} \con
\item When $z>1$,   a  non-zero fraction of the nodes belongs to the unique giant percolating component and the thermodynamics  of  statistical models (like the Ising model) is non-trivial, i.e. a phase transition may be present.
\item Typical loops have a length of $O(\log(N))$: the probability that a node belongs to a loop of length $d$ is equal  to $z^d/(2N)+ O(N^{-2})$ for large $N$.
\item If we call $z_i$ the number of nearest nodes of the node $i$, the variable $z_i$ has a Poisson distribution with average $z$. For this reason, this ensemble is also called a {\it Poisson Bethe Lattice}.
\end{itemize}

We are interested in the following spin glass Hamiltonian:
\begin{equation}
H_J(\{\sigma\})=\sum_{i,k} J_{i,k}\sigma_i \sigma_k\, , \label{H}
\end{equation}
where the sum runs over all edges of the graph, the $\sigma$'s are Ising variables defined on the nodes of the graph, and the $J$'s are randomly symmetrically distributed  quenched variables defined on the edges of the graph:  the most common   cases  are a Gaussian distribution with zero average or  a bimodal distribution ($J=\pm c$ with equal probability). In the following, we consider systems where the probability distribution of the $J$ goes fast to zero at infinity, (for example it has a uniformly bounded kyrtosis).

We are interested in computing the free energy density 
\begin{equation}
F\equiv \lim_{N\to\infty}-{\log(Z^N_J(\beta))\over \beta N}\, ,
\end{equation}
where $Z^N_J(\beta)$ is the $J$-dependent partition function on an $N$ spins system. It can be shown that this limit does not depend on $J$ with probability one: the problem is well posed. We have not indicated in an explicit way the dependence of  the free energy on $\beta$, $z$ and of the probability distribution of the $J$.

In the limit $z\to \infty$, keeping $z\langle J^2 \rangle=1$, the free energy becomes independent of the probability distribution of the $J$ and we obtain the same free energy of the SK model \cite{GT,SEN,PANK}. The Hamiltonian of the SK model has the same form of equation (\ref{H}), however, in the SK model the sum runs over all pairs of indices (we are on the complete graph) and the couplings $J$  have a variance equal to $1/N$.

\section{The  cavity approach and asymptotic Gibbs measures}
In this section, I will summarize our present understanding of the cavity approach,  both in the case where there is a unique equilibrium state and in the more  difficult case where multiple (infinite) equilibrium states are present and they are described by asymptotic Gibbs measures.

\subsection{The naive cavity equations}
Let us assume that there is only one equilibrium state. In this case, we can write the naive  cavity equations: they are well-known \cite{MP,FL} so that I would only quote the main results bypassing the physical heuristic motivations. 

 In the  naive  cavity approach, a crucial quantity is the probability distribution of the cavity magnetization $m_i$: $m_i$ is  equal to  $\langle \sigma_i \rangle$ after removing from the Hamiltonian  one among the spins that are nearest  neighbors of the spin $\sigma_i$. In the same spirit we can consider  the effective cavity field $h_i$, defined by $m_i=\tanh(\beta h_i)$.

 The equations for the cavity field distribution $P(h)$ are the following
\begin{eqnarray}
 P(h)=\overline{\delta\left(h -U_K(\vJ,\vh)\right)}\,  , \ \ \ U_K(\vJ,\vh) \equiv \sum_{i=1,K} u(J(i),h(i)) \,, \nonumber  \\ 
 u(J,h)=\frac{1}{\beta} \mbox{atanh}(\tanh (\beta J) \tanh(\beta h)) \, ,\label{NCE}
 \end{eqnarray}
where $K$ is a Poisson random number with average $z$,  the $J(i)$ are random couplings, the effective fields $h(i)$  are randomly distributed variables\footnote{I use the notation $h(i)$ and not $h_i$ in order to stress that here $i$ is not a node of the original graph. Cavity equations can also be written on a given graph, usually under the name of belief propagation equations. I will not discuss  here this approach.}  with the probability $P(h)$ , $\vJ$ and $\vh$ denote respectively the $K$-component vector of the variables $J$ and of the variables $h$. Finally the {\sl over-line} denotes the average over all the random quantities in the r.h.s. of the previous equation.   

Equivalently we could say that the random variable $h'$, defined as
\begin{equation}
h'\equiv U_K(\vJ,\vh) \, ,
\end{equation} 
have the same probability distribution of the $h$'s at the r.h.s., i.e. $P(h)$.

A crucial observation is the following. The cavity equations can be obtained by a variational principle:
\begin{eqnarray}
F_{naive}=\max_{P(h)}F_{naive}[P]\,, \\
F_{naive}[P]\equiv-\beta^{-1}\left[
% -\frac{z}{2}\log(\cosh(\beta J))+
 \overline{\log( \Delta ^{node}_K(\vJ,\vh))} -\frac{z}{2}\, \overline{\log(\Delta ^{edge}(J,h(1),h(2))})\right] \,. \label{NFE}\nonumber
\end{eqnarray}
All the $h$'s are extracted according to the probability $P(h)$ and  $K$ is (as before) a  Poisson random number with average $z$.

 The function $\log(\Delta ^{node}_K(\vJ,\vh))$ is equal to $-\beta$  times the average increase in free energy when adding to the Bethe lattice a spin that is connected  to  $K$ neighbours characterized by the field $h_i$ with interactions  $J_i$. It is given by  
 \begin{equation}
\Delta ^{node}_K(\vJ,\vh))=\sum_{\sigma,\sigma(1), \cdots ,\sigma(K)}P_1(\sigma(1)|h(1)) \cdots P_K(\sigma(K)|h(K))
\exp\left(-\beta\sigma\sum_{i=1,K}J(i)\sigma(i)\right) \,,
\end{equation}
where \begin{equation}
P_i(\sigma(i)|h(i)=1+\tanh(\beta h(i)) \sigma(i) \,.
\end{equation}
A detailed computation gives
\begin{equation}
\Delta ^{node}_K(\vJ,\vh))=2(\cosh( \beta U_K(\vJ,\vh))) \prod_{i=1,K}\frac{\cosh(\beta J(i))}{\cosh(\beta  u(J(i),h(i)))}\,.
\end{equation}
The function $\log(\Delta^{edge}(J,h(1),h(2))$ is equal to $-\beta$  times the average increase in free energy when adding to the Bethe lattice an edge with interaction $J$ between two nodes  characterized by the fields $h(1)$ and $h(2)$. It is given by  \begin{equation}
\Delta ^{edge}(J,h(1),h(2))=\sum_{\sigma(1), \sigma(2)}P_1(\sigma(1)|h(1)) P_2(\sigma(2)|h(2))
\exp\left(-\beta J \sigma(1)\sigma(2)\right)\, .
\end{equation}
A detailed computation gives
\begin{equation}
\Delta ^{edge}(J,h(1),h(2))=\cosh(\beta J) \big(1+\tanh(\beta J)\tanh(\beta h(1))\tanh(\beta h(2))\big)\, .
\end{equation}

I will not present in details the heuristic arguments \cite{MP} that lead to the free energy (\ref{NFE}). Here I would like to stress  that the function $P(h)$ that maximizes the naive free energy $F_{naive}[P]$    (\ref{NFE}) satisfies the  naive cavity equation  (\ref{NCE}). In this way the cavity equation acquires a variational meaning: they can formally be written as $\delta F[P]/\delta P=0$\,.

In the high-temperature region, there is only one equilibrium state and the true free energy is given by $F_{naive}$. In the low-temperature region (for $z>1$) multiple states exist: in this region  $F>F_{naive}$, a rigorous upper bound established by Franz and Leone \cite{FL}.

\subsection {Dealing with multiple equilibria}
 Let now assume heuristically that for a given large system we have many equilibria states\footnote
 {We are speaking of many equilibrium states for a finite system, while many equilibrium states should be present only in the infinite volume limit \cite{CINQUE}. However the considerations we present have only a heuristic value.} that are labeled by an index $\alpha$ running from 1 to $\infty$\,.
  Each state enters in the Gibbs decomposition with a weight $w_{\alpha}$, where $\sum_{\alpha=1,\infty} w_{\alpha}=1$. The effective field in each site  depends on the state $\alpha$.
 If we write heuristically the variation of the free energy when we add a new node, we have to compute  the variation of the partition function in each of the states $\alpha$, i.e. $\Delta ^{node}_{K,\alpha}$;  this quantity has the same expression as before\footnote{
 As before $K$ is a random Poisson variable with average $z$.}:
 \begin{equation}
\Delta ^{node}_{K,\alpha}=\Delta ^{node}_K(\vJ,\vh_\alpha) \,.
\end{equation}
 Here $\vh_\alpha$ is the vector $h_\alpha(i)$ for $i=1,K$. The total variation of the partition function will be given by
\begin{equation}
 \sum_\alpha w_\alpha\Delta ^{node}_{K,\alpha}=   \sum_\alpha w_\alpha \Delta ^{node}_K(\vJ,\vh_\alpha)\, .
\end{equation}
 A similar expression holds for the variation of the partition function when we add a new edge.

This expression depends on  the set of all the $w_\alpha$ and of all the $\vh_\alpha$ for all possible values of $\alpha$ and $K$. Let us define the descriptor $\cD$ which gives all the relevant information on the structure of the states. The crucial information is the values of the variables $w_\alpha$ and the fields $\vh_\alpha$; therefore we  define $\cD$ as
 \begin{equation}
\cD\equiv\{\{w_\alpha\},P({\cal H})\} . \label{DES}
\end{equation}
Here we have defined 
\begin{equation}
{\cal H}(i)\equiv \{h(i)_\alpha\}\,,
\end{equation}
i.e. 
the set of the  $h(i)_\alpha$ at fixed $i$ for all $\alpha$: this is an infinite vector as far as $\alpha$   runs up to infinity.  We know \cite{PanBethe} that  the ${\cal H}(i)$ are i.i.d.  variables. Therefore $P({\cal H})$ contains the information needed to generate all fields $\vh_\alpha$: for each $i$ ($i=1,K$)  the field ${\cal H}(i)$ is extracted randomly with the probability distribution $P({\cal H})$.

In the random case,   the descriptor $\cD$ is not unique: it will be different in different large systems; in other words, also in the thermodynamic limit the $w$'s change from large system to large system. The characterisation of the structure of equilibrium states is given by the probability distribution $\cP[\cD]$, that plays a crucial role in the theory. 

At the end of various heuristic arguments one  defines a free energy $\cF[\cP[\cD]]$ that is a functional of $\cP[\cD]$:
 \begin{equation}
\cF[\cP[\cD]] \equiv-\beta^{-1}\left[ \overline{\log\left( \sum_\alpha w_\alpha\Delta ^{node}_K(\vJ,\vh_\alpha)\right)} -{z\over2}\overline{\log\left(\sum_\alpha w_\alpha\Delta ^{edge}(J,h_\alpha(1),h_\alpha(2))\right)}\right] \, . \label{FE}
\end{equation}

The previous formula is quite complex: also neglecting the weights $w$'s, $\cP[\cD]$ a probability distribution over an infinite vector ($\alpha$ runs up to infinity). 
This formula is the natural extension of the Bethe expression for the free energy to the case where multiple states are present. It may be derived in many ways.  The simplest argument is based on cavity \cite{MP}. 
\begin{itemize} \con
\item We assume that the average value of the quantity $\Delta F_N$,  defined as the difference of the total free energy of $N+1$ and $N$ spins, coincides with the free energy density in the limit where $N$ goes to infinity.
\item When we add a spin, we  add an extra node. In this way the total number of links  increases by $z$ while the total of links should increase only by $z/2$ (in the average); in order to get the correct result we have to subtract the contribution of the extra $z/2$ links. Finally  $\Delta F_N$ has two contributions: the first coming from adding a site and the second from erasing $z/2$ links (this is the origin of  the negative  $-z/2$ factor).
\end{itemize} 

There are other  heuristic arguments \cite{MP,Parisi2003,Parisi2013} that can be used to derive the previous equation, but they are not relevant here: we will see later that the properties of the functional $\cF[\cP[\cD]]$ are deeply related to the value of the exact free energy \cite{PanBethe}.

\subsection{On descriptors or asymptotic Gibbs measures}

Descriptors are related asymptotic Gibbs measures \cite{PANEX}. The reasons is the following: in the original papers on RSB it is was proposed that the Gibbs states of a finite large system can be decomposed into  many  pure equilibria states states: we could write 
\begin{equation}
\langle \cdot \rangle_G=\sum_\alpha w_\alpha\langle \cdot \rangle_\alpha \,,
\end{equation}
 where $\langle \cdot \rangle_G$ denotes the average in the Gibbs state and the states labeled by $\alpha$ are pure clustering  states \cite{CINQUE,ParisiDescriptor}. We can define the standard overlap among these states 
 \begin{equation}
q_{\alpha,\gamma}=N^{-1}\sum_{i=1,N}\langle \sigma_i\rangle_\alpha \langle \sigma_i\rangle_\gamma \,.
\end{equation}
 We can also define generalized overlaps among these states 
 \begin{equation}
q^A_{\alpha,\gamma}=N^{-1}\sum_{i=1,N}\langle A_i\rangle_\alpha \langle A_i\rangle_\gamma \,,
\end{equation}
where $A_i$ are  local quantities, i.e. functions of only the spins that stays at a finite distance from the site $i$.
 
 To each instance of the problem (e.g. a choice of the $J$'s) we can associate to the Gibbs states a global descriptor, i.e. the weights of the states and their mutual generalized overlaps. In this language, a descriptor is given
 by
 \begin{equation}
\cD\equiv \{\{w_\alpha\}\{q^A_{\alpha,\gamma}\}\}\,.
\end{equation}
The properties of the Gibbs state and its decomposition into pure states depends on the  instance. For each $N$ we can define a probability $P_N(\cD)$ and $P(\cD)=\lim_{N\to\infty} P_N(\cD)$.

This is the standard heuristic treatment.  We can improve it in the following way.
 Instead of considering the overlaps of local quantities we can introduce local fields $h^\alpha_i$, that  are the $\alpha$-dependent fields acting on the site $i$. Once we know their distribution (given by a function  $P({\cal H})$) we can easily compute  the overlaps. For example in the case of the standard overlap we get 
\begin{equation}
q_{\alpha,\gamma}=N^{-1}\sum_{i=1,N}\tanh(\beta h_\alpha(i))\tanh(\beta h_\gamma(i)) \,.
\end{equation}

Nowadays, there is a simple and rigorous approach that avoids the introduction of the decomposition of the finite volume Gibbs state into pure states and it is based on the following construction.

 For each instance of the system
we define the set $\Sigma_N$ equal to $\{\sigma_a(i)\}$, where $i=1,N$ and  the index $a$ labels statistically independent configurations of  system extracted from the $J$ dependent Gibbs measure; the index $a$ runs from 1 to infinity. In physical language, we consider an infinite number of replicas of the same system. For a given instance the probability of $\Sigma_N$ depends on the couplings and on the topology of the lattice (that we denote collectively by $ J$): we call it 
\begin{equation}
P_J(\Sigma_N) \,.
\end{equation}
This distribution factorizes in the product of an infinite number of Gibbs distributions, each for a given replica $a$.
We are finally interested in the quantity 
  \begin{equation}
P_N(\Sigma_N) =\overline{P_J(\Sigma_N) }\,,
\end{equation}
where the average is done over all the possible instances of the problem. 

This probability distribution is called the Gibbs measure on the replica space. It does not factorize into the product of independent distributions: the different replicas are correlated as the effect of the average over the random Hamiltonian. 

We can now consider that limit $N\to\infty$ of $P_N(\Sigma_N)$. This limit exists (al least by subsequences); we call it the asymptotic Gibbs measure (i.e. $P_\infty(\Sigma_\infty)$).
We can now bypass all the discussions on the state decomposition of the finite system and consider directly the properties of the asymptotic Gibbs measure. Using the Aldous-Hoover representation it can be shown \cite{PANEX} that we can introduce the descriptors $\cD$ as in eq. (\ref{DES}) and the  probability distribution $\cP[\cD]$. 

Which are the properties of the Asymptotic Gibbs measures? Let me present the most important one (proved or conjectured). 
\subsubsection{Stochastic stability}
Stochastic stability  (i.e the Ghirlanda Guerra identities \cite{GuerraS,GuerraSS} and their later generalizations \cite{AiCon,ParisiSS}) is a crucial property that has very far-reaching consequences, among them ultrametricity \cite{PANU}. It is not worthwhile to define  exactly stochastic stability: heuristically it states that the properties of the system are stable with respect to random stochastic perturbations.  It is not rigorously  known if our  original problem, i.e. spins glass on a Bethe lattice, is stochastically stable; however, we can transform it into a stochastically stable system by adding an arbitrarily small perturbation in the Hamiltonian.
In the following, for sake of simplicity, we  assume that  our system is stochastically stable and we do not mention this arbitrarily small perturbation in the Hamiltonian.

Stochastic stability induces an infinite set of identities on the probability distribution of the overlaps. For example, we can define the probability of the overlap of two different replicas $a,b$ with $a\ne b$:
\begin{equation}
P_{a,b}(q_{a,b})\equiv P(q).
\end{equation}
This probability distribution does not depend on $a$ and $b$: therefore we can suppress the reference to these indices.

In a similar vein, we can consider four different replicas($a$, $b$, $c$ and $d$) and the conjoint probability distribution of the overlaps $q_{a,b}$ and $q_{c,d}$  that we call $P_{a,b;c,d}(q_{a,b},q_{c,d})$. Stochastic stability implies
\begin{equation}
P_{a,b;c,d}(q_{a,b},q_{c,d})= \frac13 P(q_{a,b}) \delta(q_{a,b}-q_{c,d})+\frac23 P(q_{a,b}) P(q_{c,d}). 
\end{equation}

The most spectacular consequence of stochastic stability is ultrametricity: after many partial results it has been has finally  proved by Panchenko that stochastic stability implies that the set of states is ultrametric \cite{PANU}. Using the same notation as before, ultrametricity states that
\begin{equation}
P_{a,b,c}(q_{a,b},q_{b,c}q_{c,a})=0 \ \ \ \mbox{if} \ \ q_{b,c}<\min (q_{a,b},q_{a,c})\,.
\end{equation}
Using the distance $d_{a,b}=1-q_{a,b}$, the previous formula implies that this distance is ultrametric:
\begin{equation}
d_{b,c}\le \max (d_{a,b},d_{a,c})\,. 
\end{equation}

Ultrametricity and stochastic stability have remarkable consequences: for example, all the conjoint probability distributions of the overlaps of an arbitrary number of replicas can be computed from the knowledge of the function $P(q)$. Moreover, we can associate to the set of states a taxonomic tree ($\cT$).
Each  leaf is labeled by $\alpha$ and it carries  a weight $w_\alpha$  
This tree must have an infinite number of branches and leaves (excluding the replica symmetric case, where the tree has only one leaf).  

The tree $\cT$ changes from instance  to instance of the system; for a given model there is a probability distribution on the space of trees. In the next section, we shall discuss in details the properties of this probability distribution. 

\subsubsection{The nature of the order parameters: the issue of  reproducibility}
Usually, when multiple equilibria are present, the different equilibrium states are characterized by an order parameter, e.g. the magnetization. Inside a given equilibrium state intensive quantities do not fluctuate.
What happens in the spin glassy phase? There is one conjectured property, reproducibility, that answers to this question \cite{MV}.

Reproducibility is a generalization of overlap equivalence  \cite{PRT}: it states that all the mutual information about  many  equilibrium
configurations is encoded in their mutual  overlaps. In other words, according
to this principle,  any possible local definition overlap or multioverlaps (see below for the definition of multioverlaps) should not give
information additional to that coming from the usual overlap. In other words, any intensive function of the  three (or more) configurations  does not fluctuate at fixed value of the overlaps: its value is reproducible when we change the equilibrium configurations.

A partial positive proof of reproducibility come from the  Panchenko's synchronization theorem  \cite{MS1} based on stochastic stability. Roughly speaking this theorem  states that  the generalized overlap of two states $\alpha$ and $\gamma$ ($q^A_{\alpha,\gamma}$) is a function of the standard ovelap  ($q_{\alpha,\gamma}$):
\begin{equation}
q^A_{\alpha,\gamma}=L^A(q_{\alpha,\gamma}) \,,
\end{equation}
where $L^A(q)$ is a function that depends on $A$ and on all the parameters of the problems.

However, this very nice theorem does not prove reproducibility. Things become more complex when we introduce multioveraps. Let me present the simplest example of multioverlaps.

 We have three states $\alpha$, $\beta$ and $\gamma$. We can define the standard overlaps (i.e. $q_{\alpha_,\gamma}$)  and the multioverlap $q^{(3)}_{\alpha,\beta,\gamma}$ as 
  \begin{equation}
q_{\alpha_,\gamma}=\lim_{N\to\infty}N^{-1}\sum_{i=1,N} m_\alpha(i) m_\gamma(i) \, \ \ \ 
q^{(3)}_{\alpha,\beta,\gamma}=\lim_{N\to\infty}N^{-1}\sum_{i=1,N} 
m_\alpha(i)\, m_\beta(i)\, m_\gamma(i) \,,
\end{equation}
the quantities  $m_\gamma(i)\equiv \langle \sigma(i)\rangle_\gamma$ being the magnetizations at the site $i$.
Equivalently  we could write
\begin{equation}
q_{\alpha_,\gamma}=\langle \tanh(\beta h_\alpha)\tanh(\beta h_\gamma) \rangle_{\cal H}  \, \ \ \ 
q^{(3)}_{\alpha,\beta,\gamma}=\langle \tanh(\beta h_\alpha)\tanh(\beta h_\beta)\tanh(\beta h_\gamma) \rangle_{\cal H}\,,
\end{equation}
where $\langle \cdot \rangle_\cH$ denotes the average  done with respect to the distribution $\cP(\cH)$.

For reproducible asymptotic Gibbs measures   the values of the multioverlaps do not fluctuate at fixed values of the overlaps.  If reproducibility holds multioverlaps are  functions of the standard  two indices overlaps $q_{\alpha_,\gamma}$, $q_{\beta,\gamma}$, $q_{\alpha_,\beta}$\,, i.e., there is a function $Q^{(3) }$, such that 
\begin{equation}
q^{(3)}_{\alpha,\beta,\gamma}=Q^{(3) }(q_{\alpha_,\gamma}, q_{\beta,\gamma}, q_{\alpha_,\beta}) \,,
\end{equation}
and the values of the standard overlaps determine the multioverlaps. 

Using the language of \cite{MV} reproducibility states that  all the multioverlaps are reproducible functions of the overlaps, hence the name {\sl reproducible}. The fluctuation of the multioverlaps in the ensemble at fixed overlaps must go to zero in the infinite volume limit.

Can reproducibility  be proved? A  theorem by Panchenko \cite{PanBethe} states that a stochastically stable asymptotic Gibbs measure that satisfies  the cavity equations (to be defined in the appendix)  must be reproducible if the overlap $q$ may take only $r+1$ possible values (i.e. $r$ steps RSB). 

This theorem does not forbid the existence of stochastically stable {\sl non-reproducible} distributions (satisfying the cavity equations) in the case of continuous replica symmetry breaking.  These  distributions, provided that they exist, cannot be obtained as the limit $r\to\infty$ of stochastic stable distributions (satisfying the cavity equations). Personally, I would extremely surprised by the existence of such  non-reproducible distribution. However, their non-existence for the moment is an unproven conjecture. In the rest of this paper, I will consider only reproducible distributions.

 \section{Computing the free energy}
 This paper is based on the following   wonderful theorem \cite{FL}: 
\begin{equation}
F\ge F_{R}\equiv\max_{\cP_{R}[\cD]}\cF[\cP_{R}[\cD]] \, , \label{FINAL}
\end{equation}
where the probability distribution $\cP_{R}[\cD]$ is a stochastically stable distribution
 concentrated on the set of  reproducible descriptors that are defined above.  Indeed if we are interested in the evaluation of the free energy   it has been proved \cite{GuerraSS} that we can limit ourselves to the study of stochastically stable probability distributions.
 
 This theorem is valid for the Poisson Bethe lattices. A similar theorem exists for models defined on Random Regular Graphs, with a slightly different expression for the free energy \cite{RRG}. 
 
We shall discuss in details the properties of $\cP_{R}[\cD]$ and how to compute $F_{R}$ in the rest of the paper. However, before going on, I would like to mention a very nice theorem \cite{PanBethe}: it shows that \begin{equation}
F=F_C\equiv\max_{\cP_C[\cD]}\cF[\cP_C[\cD]]\ , 
\end{equation}
where  the probability distribution $\cP_C[\cD]$ satisfy the cavity equations \cite{PANEX}. I have already mentioned that it is conjectured  that these distributions (if they are also stochastically stable) reduce to the closure of the set of  stochastically stable reproducible distribution. Unfortunately such theorem (in spite of the partial results in \cite{PanBethe}) is not {\sl yet} proved. For the time being the relation
\begin{equation}
F_{R}=F_C
\end{equation}
remains as an unproved conjecture, in spite of the  general expectation that  the equality $F=F_{R}$ is correct and that the computation of $F_{R}$ solves the problem. 

We can now address the problem of producing explicit formulae for $\cP_{R}[\cD]$ and of using them to compute the free energy $F_R$.
\subsection{The distribution of the weights} 
Our first task is to discuss the distribution of the weights of the states in a stochastically stable ensemble.
In this case, ultrametricity holds and the states are organized in a taxonomic tree.  This point is widely discussed in the literature in the case of the SK model; there are no changes when we go to Bethe lattice model, so we will only summarize the main results.

One can assign  to each branching point of the tree of the states a variable $x$ that may ranges from zero to $x_M<1$: the root is at $x=0$ and the leaves (the states) are at $x=x_M$.

 Roughly speaking the trees may be of two kinds \cite{MPV,ParisiTree}:
\begin{itemize} \con
\item The trees may have points only at a discrete number ($r$) of levels. In this case, the probability distribution of the weights is automatically fixed by the position of the $r$ levels ($x_a$, $a=1,r$); sometimes $r$ is called the number of steps. We shall see later explicit formulae for this probability distribution.
\item The trees may have  branching points at any level in the interval $[x_m,x_M]$: these trees are usually called continuous trees. In this case the probability distribution of the trees is essentially unique; it can be constructed in a rigorous way or using the Ruelle cascade approach \cite{Ruelle} (from the root of the tree toward the leaves) or using the Bolthausen-Sznitman  coalescent approach \cite{Bolt} (from the leaves toward the root of the tree). The probability distribution over continuous trees may also be constructed in a  weak sense as the limit of $r$ going to infinity of the probability over discrete trees when $r\to\infty$.
\end{itemize}

In the rest of the paper, we will speak of the tree of the states referring to this construction. In both cases, we know the probability distribution of the trees, including the associated weights of the leaves (i.e. $w_\alpha$). The explicit formulae will be written below. 

We only notice that in the case of a discrete number ($r$) of levels
instead of labeling the states by an integer $\alpha$ we label them a multi-index
\begin{equation}
\alpha\equiv\{\alpha_1, \dots \alpha_r\} \, ,
\end{equation}
where the new indices $\alpha$'s range from $1$ to $\infty$. This last set has the same cardinality of the integers (Cantor's construction): the introduction of the multi-index is just a convenient relabelling. Using the language of \cite{MPV,ParisiTree}, in the case $r=3$ we could say  that states are like plant names in the taxonomic classification, that are labeled by the {\sl family} ($\alpha_1$), the {\sl genera}  ($\alpha_2$) inside the family and the {\sl species} ($\alpha_3$). More precisely all the states that have the same pair ($\alpha_1,\alpha_2$) belong to the same cluster (that we can label by $\alpha_1,\alpha_2$) and all the states that have the same $\alpha_1$ belong to the same supercluster. 

If we define the weight of a node as the sum of the weights of its descendants, there are recursive rules that give the weight of the suns as a function of the weights of the fathers, that can be applied recursively starting from the root that has weight 1. Having tamed the distribution of the weights, our next task is to control the conjoint distribution of the weights and the field.
\begin{figure}
\begin{center}\includegraphics[width=.8\linewidth]{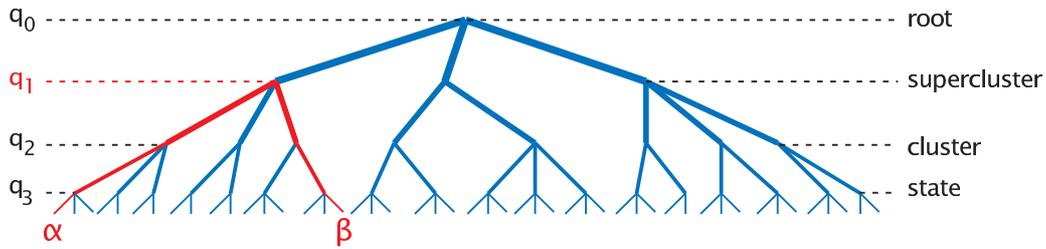}\end{center}
\caption{Taxonomic structure of the
tree of states in a simplified example with $r=3$, taken from \cite{PRY}.  
%Notice that the overlap between states $\alpha$ and $\beta$ is  $q_{\alpha\beta}=q_1$.
\label{fig:arbol}}
\end{figure}

\subsection{Some examples of Reproducible Descriptors} 

The construction of reproducible descriptors \cite{PanBethe} we  are going to present seems to be rather complex, however, it is the simplest one that satisfies all the needed theoretical constraints. In the following, we will explicitly construct   reproducible descriptors  when the tree of states has only $r$ levels.

We  can consider of sthocastic stable probability distributions $\cP[\cD]$ that factorize into a simple product \cite{MP,PanBethe}:
\begin{equation}
\cP[\cD]=Q[\{w\}] P[\cH].
\end{equation}
 The crucial problem is the choice of  $P[\cH]$. We  shall see below how to construct the reproducible descriptors for finite $r$.

% In the continuous case   we  will  define {\sl reproducible distribution} as the  limit $r\to\infty$ of the reproducible distributions for finite $r$\footnote{ This is one of the obstructions that block a possible proof of $F=F_{R}$.}. 

Here the probability $P(\cH)$ can be written in the following constructive way. Let us call $\Theta$ a given continuous space and let us consider a probability measure $d\mu(\theta)$ on $\Theta$. 
 We introduce a function $g $ of the variables $\{\theta_0,\theta_1,\dots, \theta_r\}$. We finally put
 \begin{equation}
h_{\alpha_1, \dots \alpha_r}(i) =g(\theta_0(i),\theta_{\alpha_1}(i), \theta_{\alpha_1,\alpha_2} (i),\dots, \theta_{\alpha_1,\alpha_2,\dots\alpha_r}(i)) \label{WOW} \, .
%h_{\alpha_1, \dots \alpha_r}(i) =g(\theta_0(i),\theta_{\alpha_1}, \theta_{\alpha_1,\alpha_2} \dots, \theta_{\alpha_1,\alpha_2,\dots\theta_r}) \label{WOW} \, .
\end{equation}
Here the $\theta$'s are i.i.d. random variables extracted with the probability measure $d\mu(\theta)$.
The fields $\cH_i$ (that are i.i.d. random infinitely dimensional variables) are generated by taking in the previous formula the $\theta$'s as i.i.d. random variables. The measure $d\mu(\theta)$ and the function $g$ may be a convenient compact way to describe the probability distribution $P(\cH)$. 

Formula (\ref{WOW}) may look impenetrable, so I will discuss   it in some simple cases in the next subsections. However before discussing the details let me present some general considerations.
\begin{itemize}
\item
Why this probability distribution is reproducible? Let us consider a very explicit example with $r=3$. We consider the tree states 
\begin{equation}
\alpha=\{\alpha_1,\alpha_2, \alpha_3\}\,, \ \ \ \beta=\{\alpha_1,\beta_2, \beta_3\}\,,      \ \  \gamma =\{\alpha_1,\beta_2, \gamma_3\} \,,
\end{equation}
where all the indices with different names are different. 

A simple computation shows that the value of the standard overlaps between the states does not depend on the values of the indices, as soon as they are all different. A detailed computation \cite{PanBethe} shows that also the values of the multioverlaps do not depend on the values of the indices, hence reproducibility. Equivalently one could take equation (\ref{WOW}) as the definition of reproducibility for finite \cite{PANEX,PanBethe}.

\item The introduction of the space $\Theta$ produces a  very  redundant formulation: this redundancy will be useful when will address the crucial question of the  smoothness of the function $g$.
 In the original formulation \cite{PanBethe},  the variables $\Theta$ belong to the interval [0,1]  and have a flat distribution. This formulation is perfectly adequate  if we do not raise the issue of smoothness of the function $g$. 
 
Indeed we know that the Peano curve  maps the interval [0,1] into the two-dimensional square in such a way that every point of the square belongs to that curve; however,  the Peano curve is a continuous  non-differentiable curve:  it is Holder 1/2. We cannot construct a  differentiable Peano curve. We can generalize the Peano construction: for example,  any random object in a nice enough space (e.g. a complete
separable metric space) can always be generated by a function from [0,1] to
that space. Basically, any randomness can always be encoded by a uniform
random variable on [0,1]. This follows from  the  Borel isomorphism theorem\footnote{I am grateful to Dmitry Panchenko for clarifying this point to me.}. 

However for practical purposes (e.g. numerical approximations) dealing with non-differentiable function is a mess and it is much simpler to work with sufficiently smooth functions. Representing a $D$ dimensional smooth manifold in $M$ dimensions with a function from the interval $0-1$ to that space is not the most  popular and useful choice.

In general, the smoothness properties of the function $g$ depend on the choice of the dimension $D$ for the space $\Theta$. It would be  very interesting  to have more information on the relations among $D$ and the smoothness properties of the function $g$. This would give the information of how the set of the probability distributions can be approximated by a finite-dimensional manifold: these results may also be useful for constructing approximations.

\item I would like to add an extra remark. In this model, the numbers of neighbors of a given spin is a fluctuating Poisson random variable. The probability distribution of the fields depend on the number of neighbors, therefore the total probability will receive the contributions from different probability distributions.  This is an extra complication that may be removed in two ways:
 \begin{itemize}
 \item We explicitly introduce different probabilities for spins with a different number of neighbors.
 \item We change the model and we consider a Bethe lattice with fixed number of neighbors, i.d. the random regular graph.
 \end{itemize}
 The second approach is certainly simpler to implement numerically.
\end{itemize}

\subsubsection{The case $r=0$: replica symmetry holds}
 We have only one state (no $\alpha$'s) and therefore only $\theta_0$. Here we have 
\begin{equation}
h(i)=g(\theta_0(i))\,, \ \ \ P(h)=\int d\mu(\theta_0)\delta (h-g(\theta_0))\, ,
\end{equation}
where $i$ labels the different sites. In this case the $h(i)$ are i.i.d. variables with a probability distribution $P(h)$, i.e. the  result of the naive cavity approach.

It is clear that the definitions of the space $\Theta$ and of the measure $\mu$ are arbitrary: we could have taken the variable $\theta$ as a flat distributed continuous variable in the interval [0,1]. Here the introduction of the space $\Theta$ is  useless.
The one dimensional function $g(\theta)$  can be constructed in a simple way: we define the cumulative function
\begin{equation}
C(h)=\int_{-\infty}^h d h' P(h')\, .
\end{equation}

If the function $P(h)$ is smooth (e.g. it contains no delta functions) the function $C(h)$ takes all the  possible values in the interval $[0,1]$. The natural choice of  the function $g(\theta)$, $\theta$ being in the interval [0,1], is given by the condition
\begin{equation}
g(C(h))=h\, .
\end{equation}
With this choice $g(\theta)$ in a monotonous function. The motonicity condition determines the function $g$ in an unique way: there are many different non-monotonous $g$ functions that give the same $P(h)$.

\subsubsection{The case $r=1$: 1 step RSB}
 Here $\alpha\equiv \alpha_1$ and the function $g(\theta_0, \theta_1$) depends on two variables. 
 \begin{equation}
h_\alpha(i)=g(\theta_0(i),\theta_\alpha(i))\, .
\end{equation}At fixed $i$ for a given choice of  $\theta_0(i)$, the variables $h(i)_\alpha$ ($\alpha=1\dots\infty$) are i.i.d. variables with the probability distribution
\begin{equation}
P_i(h)=P(h|\theta_0(i))\equiv\int d\mu(\theta)\, \delta (h-g(\theta_0(i),\theta))\, .
\end{equation}
In each point $i$ we have a different probability distribution (i.e. $P_i(h)$)  that is parametrized by the random variable $\theta_0(i)$. 

The values of the fields for the same value of $\alpha$ at different points $i$ and $i'$ for $i\ne i'$ (i.e. $h_\alpha(i)$ and $h_\alpha(i')$)  are uncorrelated because the varaibles $\theta_0(i)$ and $\theta_\alpha(i)$ are respectively uncorrelated with $\theta_0(i')$ and $\theta_\alpha(i')$. 
%\footnote
%{A  non-reproducible probability distribution can be constructed as follows. The probability distribution of the  variables $h_\alpha(i)$  depends on both $\alpha$ and $i$: these variables are extracted with a probability distribution $P_\alpha(i)(h)$. The values of the fields for the same value of $\alpha$ at different points $i$ and $i'$ are now correlated. Fortunately it was proved in \cite{PanBethe} that this much more complex non-reproducible construction is forbidden for finite $r$ under a technical hypothesis, i.e.  the overlap $q_{\alpha, \gamma}$ may take only $r$ different values.  }.

In this case, we can avoid the introduction of the function $g$ and of the  $\theta$'s: we could  describe the situation by defining the probability $\cP[P]$, i.e. the probability that  at a given point the probability of the $h_\alpha$ is $P$. This description is done in terms of the probability function over probability functions\footnote{In  \cite{MP} the authors introduced a  population of populations in order to formulate the problem in such a way that  numerical computations are possible.}. Such a  functional probability distribution is a somewhat esoteric object, so it may be  more convenient to introduce a function of two variable at its place, losing the uniqueness of the description.

Let us now come to the issue of the smoothness of the  function $g$. Let us assume for simplicity that there is natural (differentiable) parametrization of $P_i(h)$ in terms of a vector $\vec{v}$ that is flatly distributed on the $D$-dimensional cube, i.e. $P(h|\vec{v})$. In this case for each $\vec{v}$ we can introduce the function
\begin{equation}
g(\vec{v},C(h|\vec{v}))=h\, \ \ \ \ \mbox{where} \ \ \ C(h|\vec{v})=\int_{-\infty}^h d h' P(h'|\vec{v}).
\end{equation}
This natural description corresponds to take the first $\theta$ in the $D$-dimensional cube and the second $\theta$ in the interval $[0,1]$. 

When we study the free energy at zero temperature, we know that the relevant probability distributions can be parametrized in terms of the surveys (\cite{MP2, MPZ}). Surveys  form a two-dimensional manifold. In this case, a two-dimensional space for the first variable would be adequate.

It is worthwhile to note that in the SK limit (i.e. $z\to\infty$) the choice $D=1$ is perfectly adequate: the distribution probability $P(h)$ can be written  as a smooth  function of a parameter $h_0$: $P(h|h_0)$. The form of the distributions $P(h_0)$ and $P(h|h_0)$ are simple and well known \cite{MV}. Later we  will find a detailed discussion of this point.

\subsubsection{The case $r \geq 2$: many steps RSB} 

In the  $r=2$ case for each site $i$ and each cluster $\alpha_1$ (i.e. for a given $i$ and $\alpha_1$), the  fields $h(i)_{\alpha_1,\alpha_2}$  ($\alpha_2=1\dots\infty$)  are i.i.d. variables: their probability distribution  is given by
\begin{equation}
P_{i,\alpha_1}(h)=\int d\mu(\theta)\delta (h-g(\theta_0(i),\theta_{\alpha_1}(i),\theta)).
\end{equation} 

States that belong to different clusters (i.e. they have different values of $\alpha_1$) have  different probability distributions. Each site is characterized by the cluster dependent probability distribution that the states of each cluster  have a given probability distribution for the fields at that point. Finally one arrives at a  probability distribution of probability distributions over probability distributions. The function $g$ is a simple way to describe and parametrize this mess, maybe putting some dirt below the carpet.

In this case we can ask a new question. Let us suppose that we can write a function $g$ that is sufficiently smooth when the three $\theta$ belong to a $D$ dimensional space. However everything would be simpler if we could write
\begin{equation}
g(\theta_0,\theta_1,\theta_2)=\hat{g}\big(\rho(\theta_0,\theta_1), \theta_2\big)\, ,
\end{equation}
where $\rho$ is a $L$ dimensional vector (with $L<2D$) and both the function $\rho$ and $g$ are sufficiently smooth. In other words we ask if the information contained in the $2D$-dimensional vector ${\theta_0,\theta_1}$ can be conveyed  in a smooth way to a variable $\rho(\theta_0,\theta_1)$ defined on a lower dimensional space.
In the best of the possible worlds, $L=D$ would be a good approximation. 

In the following we will assume that $L=D$ make sense. In  the general case of many steps we would like to know if we can write
\begin{eqnarray}
g_4(\theta_0,\theta_1,\theta_2,\theta_3,\theta_4)=\hat{g}_3\big (\rho_1(\theta_0,\theta_1), \theta_2,\theta_3,\theta_4\big)\, , \nonumber\\
\hat{g}_3(\rho_1,\theta_2,\theta_3,\theta_4)=\hat{g}_2\big(\rho_2(\rho_1,\theta_2),\theta_3,\theta_4\big)\ , \nonumber\\
\hat{g}_2(\rho_2,\theta_3,\theta_4)=\hat{g}_1\big(\rho_3(\rho_2,\theta_3),\theta_4\big)\ ,  \label{Markov}
\end{eqnarray}
where all the variables $\theta$  and $\rho$ are $D$ dimensional vectors and the values of the {\sl smooth} functions $\rho$ are  also $D$-dimensional vectors\footnote{In the first line of equation (\ref{Markov})  $\rho_1(\theta_0,\theta_1)$ is a function,
in the second line  of the same equation $\rho$ is a variable.}.

If equations (\ref{Markov}) are satisfied, the function $g$ is a {\sl Markov} function because similar conditions are true for the conjoint probability of events in a Markov chain (or process). 

If we could restrict our analysis to $D$ dimensional Markov functions (this approximation should give the free energy with a reasonable accuracy that improves with  increasing $D$), the whole computation could be greatly simplified.  It is important to note that in the SK limit, as we will see below, the previous formulae are valid with $D=L=1$, as also discussed in \cite{MV}.
 
\subsection{More details on the weights of the states}

The probability $Q[\{w\}]$ depends on $r$ parameters $x_a$  such that\begin{equation}
0<x_1<x_2<\dots <x_{r-1}<x_r \, .
\end{equation}
Fro completeness I will present a  description of  the generation of  the probability $Q[\{w\}]$ in a constructive way (mostly taken from \cite{PRY}).

I recall that in the multi-index construction we set $\alpha={\alpha_1,\dots,\alpha_r}$. The states are the leaves  of the tree of state; on this tree, the branching points of the first level are labeled by $\alpha_1$, those of the second level by $\alpha_1,\alpha_2$ and so  on.  Here we are interested   in the probability generation of the weights of the states $\{w_\alpha\}$.
To this end, we shall construct  the weights,
starting from the root  and branching out step by step down to the individual
states, i.e., the leaves of the tree.  

In this section we explain how such a construction can be implemented, starting
with the simplest case $r=1$, and then generalizing to generic $r$. We are not going to discuss how to construct the probability distribution directly in the continuous limit \cite{Ruelle,Bolt,COA}.
\subsubsection{$r=1$}\label{sec:onestep}
The  states depend only on a single index $\alpha$. The weights can be constructed in the following way.
We consider a Poisson point process with an intensity $\exp [\beta x_1 (f-f_0)]$. More precisely we extract the points $f_\alpha$ on the
line: the probability of finding a point in the interval $[f:f+\dd f]$  is parametrized by $x_1$  and it is
given by
\begin{equation} \label{eq:rhom}
\dd\rho_{x_1}(f)\equiv  \exp \bigl[\beta x_1 (f-f_0)\bigr] \dd f\, .\label{1RSB0}
 \end{equation}
We now  write
\begin{equation} w_{\alpha}=\frac{\exp(-\beta f_\alpha)}{\sum_\gamma
\exp(-\beta f_\gamma)}\, . \label{1RSB}  \end{equation}
The weights generated in this way
have the correct stochastically stable probability distribution $Q[\{w\}]$. A few comments are in order:
\begin{itemize}
\item For $x_1<1$, the construction is consistent, i.e., $\sum_\gamma
\exp(-\beta f_\gamma)<\infty $ and $\sum_\alpha w_{\alpha}=1$.
\item The parameters $f_0$ is irrelevant.
\item If we consider a Poisson point process where the $f_\alpha$ are restricted in the interval
$[-\infty,\Lambda]$ the
probability distribution of the $w_\alpha$ converges to the right one in the limit
$\Lambda\to +\infty$.
\item This probability distribution is the unique stochastically stable probability distribution on trees with only one branch point. Indeed if we define $f'_\alpha=f_\alpha+\eta_\alpha$ were the quantities $\eta_\alpha$ are i.i.d. random variables, the conditions that the variable $w'_\alpha$ have the same distribution of the original variables $w_\alpha$ implies \cite{Parisi2003,RUZ} that the distribution of the $w$'s is given by eqs. (\ref{1RSB0},\ref{1RSB}).
 \end{itemize}

\subsubsection{$r>1$}

In the case $r=2$ we can simply generalize the previous equations. We  label
the states by a pair of indices $\alpha_1$ (cluster) and $\alpha_2$ (state
within each cluster) and we define 
\begin{equation}
w_{\alpha_1}=\sum_{\alpha_2}w_{\alpha_1,\alpha_2} 
\end{equation}
or equivalently
\begin{equation}
w_{\alpha_1,\alpha_2}=w_{\alpha_1}t_{\alpha_1,\alpha_2}\, ,\ \ \ \ \sum_{\alpha_2} t_{\alpha_1,\alpha_2}=1\, .
\end{equation}
Finally, we write 
\begin{equation}
t_{\alpha_1,\alpha_2}={\exp(-f_{\alpha_1,\alpha_2})\over
\sum_{\alpha_{2}}\exp(-f_{\alpha_1,\alpha_2}) } \, .
\end{equation}
Now the $t_{\alpha_1,\alpha_2}$ are not independent, 
since they are constrained to belong to the same cluster $\alpha_1$.

The probability distribution of the $f_{\alpha_1,\alpha_2}$ can be found 
in the literature, see eq.~(14) in~\cite{MPVART}:
\begin{equation}\label{eq:Ph}
d \mathcal P{\{f\}} \propto \biggl(\prod_{\alpha_2} \dd\rho_{x_2}(f_{\alpha_1,\alpha_2})\biggr) \biggl(\sum_{\alpha_2} \exp(-\beta f_{\alpha_1,\alpha_2})\biggr)^{x_1}\, ,
\end{equation}
where $\dd\rho_{x_2}$ is a Poisson point process. 

A simple computation shows that the formulae make sense only for $x_2>x_1$. Moreover in the limit
where $x_2 \to x_1$ there is one $t_{\alpha_1,\alpha_2}$ that becomes equal to 1 while all others are going to zero.

It is important to note that the probability distribution of a single $w_{\alpha_1,\alpha_2}$ is the same as those of the $r=1$ case: it is given by the $r=1$ formula for $x_1=x_2$. The crucial difference with the $r=1$ case  is that the conjoint probability distribution of two or more $w$'s is different.  %the free energy with the same value of $\alpha_1$ and different value of  $\alpha_2$ (i.e. $f_{\alpha_1,\alpha_2}$ and  $f_{\alpha_1,\alpha'_2})$ are correlated: they have a different distribution from those with a different values of $

The cases $r>2$ can be approached in a similar way constructing the weights in a recursive way. 

\section{The continuous limit, i.e. $r\to\infty$ }

Up to now everything is clear.  We have seen that according to common lore  the upper bound for the free energy ($F_{R}$) is reached only in the limit where $r \to \infty$, the so-called continuous limit. Unfortunately, it is not easy to control what happens when $r \to \infty$. Moreover many important properties, e.g. marginal stability, are valid only in this limit: for this reason  it important to  control this limit.  At this end, I will try to  formulate the theory directly in the continuum limit and only at a later stage to introduce the approximations that may be necessary to perform an explicit computation. 

We have already remarked that is possible to introduce a measure directly on the tree where the levels are labeled by a continuous variable $x$. The construction can be done using the  observation that if we neglect states that have a weight $w_\alpha<\epsilon$, the number of branches becomes finite: the operation of cutting away the branches with weight less the $\epsilon$ is sometimes called {\sl pruning}. We can write an explicit  measure on these $\epsilon$-pruned trees \cite{Ruelle,PRY}, we can compute many of the needed properties and we can also generate these trees numerically in a constructive way\footnote{From the numerical viewpoint \cite{PRY} it may also be convenient to do the computation at finite, but large $r$, provided that the computational weight does not increase too fast with $r$.}. In the limit $\epsilon \to 0$ the measure on $\epsilon$-pruned trees converges to a measure on  infinite branched trees in a precise mathematical sense. We can also adapt the algorithm of  \cite{COA} to construct numerically a random representative of the tree of states.
   
The real problems come from the space $\Theta^r$: it is not simple to keep under control a generic function on this space when $r\to\infty$. Here we would like to point out a possible approach to deal with this problem. Unfortunately, we are not going to show in this paper that this approach works, however, we feel that also an incomplete attempt may be useful.
\subsection{The solution of the SK model}
Before going on it is convenient to recall what happens in the SK model. 
In the limit where $z$ is large, neglecting terms that go to zero when $z$ goes to $\infty$ (keeping $z\langle J^2 \rangle=1$), the Bethe lattice tends to the SK model \cite{GT}, in the sense that the free energy of the  Bethe lattice tends to the free energy of the SK model  \cite{SEN,PANK}. In this limit, we can neglect high orders in powers of $J$ in the cavity equations for the Bethe lattice.
For example  the  naive cavity equation for the SK model can be obtained  linearizing the cavity equations for the Bethe lattice:
\begin{equation}
h'_\alpha= \frac{1}{z^{1/2}} \sum_{i=1,z} J(i) m_\alpha(i)\ , \ \ \ \ m_\alpha(i)=\tanh(\beta h_\alpha(i))\, .
\end{equation}
The central limit theorem tell us that the variables  $h'_\alpha$ are  Gaussian variables with variance $q\equiv \langle m^2_\alpha \rangle$.

Let us consider two states $\alpha$ and $\gamma$  that have the last common ancestor at $x$: we will use the notation $x=\alpha \cap \gamma$. The crucial quantity is the covariance of the Gaussian variables $h'_\alpha$:
\begin{equation}
\langle h'_\alpha h'_\gamma \rangle = q(\alpha \cap \gamma)= \langle m_\alpha m_\gamma \rangle_w\, ,
\end{equation} 
where the states $\alpha$ and $\gamma$ are weighted with the weights $w$.

If we know the function $q(x)$, we have everything we need to go on. We can generate the tree of state ($\cT$), and for each tree $\cT$ we can generate the Gaussian variables $h'_\alpha$ whose covariance is given by the previous equations. We use now eq. (\ref{HCE}) and we can do the needed computations. In order to satisfy the cavity equation, we can compute
\begin{equation}
q'(x)=\langle m'_\alpha m'_\gamma \rangle_{w'}\bigr\vert_{\alpha\cap\gamma=x}\, ,\ \ \ m'_\alpha=\tanh(\beta h'_\alpha) \ ,
\end{equation}
where the various states are weighted with the new weights $w'$. The whole procedure has been implemented numerically \cite{PRY} up to the end.

It is crucial to note that the distribution probability of the $h'_\alpha$ is Gaussian. However, in the cavity equation, we are interested in the distribution of the fields after the introduction of the new weights. One finds the distribution probability of the $h'_\alpha$, weighted with the new weight $w'$ is non-Gaussian.  

The properties of the distribution probability of the $h'_\alpha$ after reweighting is crucial. These properties have been carefully studied. After some computations it can be shown that we can find    analytically  this distribution probability using the following rules \cite{Sommers,CR,MV,CONV}:
\begin{itemize}
\item We write the following  (Langevin) stochastic differential equation in the interval $[0,1]$
\begin{equation}
{dx\over dq}\,{dh_\eta(x)\over dx}=\eta(x) +\beta x \,m(x,h_\eta) \, ,
\end{equation}
where $x(q)$ is the inverse of the function $q(x)$ and  $\eta(x)$ is a white noise\footnote{If the reader does not like white noises, he can reformulate everything in terms of Brownian motions $B(x)=\int_0^x \eta(x)$: we could also use Ito stochastic calculus.}:
\begin{equation}
\overline{\eta(x)\eta(y)}= \delta (x-y) \, .
\end{equation} 
We call $h(x|x_0,h_0)_\eta$ the solution of the previous equation for $x>x_0$ with the boundary condition
\begin{equation}
h(x_0|x_0,h_0)_\eta=h_0\, .
\end{equation}
We notice that $h(x|x_0,h_0)_\eta$ depends on the values of $\eta(x')$ only in the interval $x_0<x'<x$ (a sort of Markov property).  
\item
The function $q(x)$  satisfies the self-consistency equation
\begin{equation}
q(x)=\overline{\left(m(x,h(x|0,0)_\eta)\right)^2}=\int dh P(x,h|0,0) m^2(x,h)\, ,\label q 
\end{equation}
where $P(x,h|x',h')$ (for $x>x'$) is the probability distribution of $h(x|x',h')$.
\item The function $m(x,h)$ satisfies the self-consistency equation
\begin{equation}
m(x',h')= \overline{\tanh(\beta h(1|x',h')_\eta)} =\int dh P(h,1|x',h')\tanh(\beta h(1|x',h')) \,.
\end{equation}
where the over-line denotes the average over the white noise $\eta$. It is evident that $m(1,h)=\tanh(\beta h)$. Indeed  $h(x|x_0,h_0)_\eta$ is a continuous function of $x$; consequently $\lim_{x \to x'} P(x,h|x',h')=\delta(h-h')$.
\item There is a simple iterative  procedure for solving the previous equations: for each $q(x)$ we compute the $m(x,h)$ and we use equation (\ref{q}) in an iterative way. This procedure is convergent.
\end{itemize}
It is possible to transform the previous stochastic differential equations into parabolic (or anti-parabolic) differential equations. These equations are the same as those found in the analytic solution of the SK model. We can also write a simple expression for the free energy of the SK model:
\begin{equation}
F=\max_{q} F[q]=\overline{\ln(\cosh(\beta h(1|0,0)_\eta))}-\beta^2 \int_0^1 dx{dq\over dx}\, x \,q(x)\, .
\end{equation}
The self consistent equation for the $q(x)$ (i.e. eq. (\ref{q})) can also be seen as the consequence of a variational principle with respect to the function $q(x)$ that enters in the differential equation. It can be proved  \cite{ CONV}  that the functional $-F[x(q)]$ is convex\footnote{The function $x(q)$ is the inverse of the function $x(q)$. The functional $F[x(q)]$ is concave: the concavity of this functional (proved in \cite{CONV}) is surprising and counter-intuitive, at least to me.}, so that the maximum is well defined and unique.

We now have all the tools to write down the probability distribution of the $h_\alpha$. The construction is rather simple: for each tree, 
we define the functions $\eta(x)_\alpha$ that  satisfy the   constraint
\begin{equation}
\eta(x)_\alpha=\eta(x)_\gamma \ \ \ \mbox{for}\ \ \ x<\alpha\cap\gamma \,. 
\end{equation}
The functions $\eta(x)_\alpha$ and $\eta(x)_\gamma$ are independent white noises for $x>\alpha\cup\gamma$\,.
In other words, it is natural to define the function $\eta$ on the branches of the tree. We have $\eta(x)_\alpha=\eta(x)_\gamma$ on the common part of the branches $\alpha$ and $\gamma$, while the two functions are independent and uncorrelated on the remaining part of the two branches.

We finally arrive at the simple formula:
\begin{equation}
h_\alpha=h(1|0,0)_{\eta_\alpha} \, .
\end{equation}
We can show that the $h$'s computed in this way have the correct probability distribution \cite{MV}. 

Summarizing the probability distribution of the $w$'s is concentrated on trees; for each tree, the distribution of the $h_\alpha$'s can be obtained starting from a white noise defined on the tree and integrating a stochastic differential equation.

\subsection{The minimal construction for the Bethe lattice}

Let us consider the Bethe lattice case. We make the choice   $x_a=a/r$ and we send $r$  to infinity. The quantities $U_K$ are the sum of a finite number ($K$) of variables: only in the limit where $z$ goes to infinity we  can use the central limit theorem. For finite $z$ we have to use the full distribution of the $h_\alpha$. We    have already remarked that his distribution  can be described by  a function $g(\theta_0,\theta_1,  \cdots, \theta_r)$. 

In the general case, it is very hard to control  this limit. In the following, I will make some assumptions that aim to  simplify the variational computations in the limit the $r \to\infty $ limit. 

The simplest choice is the following: in the same way as in the SK model we assume that the variables $\theta_a$ are real numbers ($\Theta$ is the real line) and the measure $d\mu(\theta_a)$ is such that the variables $\theta_a$ are Gaussian distributed with variance $1/r$. 

If we call $x=a/r$   and we set\footnote{The function $\mbox{Int}(y)$  denotes the integer part of the real number $y$.} $\theta(x)=\theta_{\mbox{Int}(rx)}$, the function $\theta(x)$ becomes a white noise  in the limit $r \to \infty$:
\begin{equation}
\overline{\theta(x)\theta(y)}= \delta (x-y) \, .
\end{equation} 

Exactly as in the SK model the root of the tree is at $x=0$ and the leaves are at $x=1$ (or at $x=x_M$). We can consider a finite $\epsilon$-pruned tree that has a finite number of branches and on each branch the white noise $\theta(x)$ is constructed in the same way as the white noise $\eta(x)$ in the SK model.

We can now assume that $h_\alpha=g[\theta_\alpha]$ for an appropriate functional $g[\theta]$, defined on the functions $\theta(x)$ in the interval $[0,1]$ (more precisely on white noises).  In this way the free energy becomes a functional of $g$, i.e. $\cF[g]$ and we must find its maximum. 

Naively speaking the problem seems to be well defined. However, we have to make some assumptions about   the functional $g$.
When we have a variational approach we may look for the solution of the variational problem restricted  on a subspace. This is a time-honored approximation: it may work quite well if the true maximum is not far from the maximum of the subspace. 

A natural idea is mimic what happens in the SK model.
A very  simple   attempt is  the following:
\begin{itemize} \con
\item We consider a function $z_\theta(x)$ that is a solution of the stochastic differential equation
\begin{equation}
{dz_\theta(x) \over dx}= A(x,z_\theta(x))\theta(x)+B(x,z_\theta(x)) \, .
\end{equation}
\item We set \begin{equation}
h_\alpha= C(z_{\theta_\alpha}(1)) \, .
\end{equation}
\end{itemize}
In this way we obtain a {\sl Markov} functional that depends only on two functions of two variables (i.e. $A$ and $B$) and one function of one variable (i.e. $C$). This  is a relative small space which can be studied in details using numerical investigation. Maybe this approximation  gives more accurate predictions for the free energy than those that are presently available ($r=1$ or 2). However the main goals of this approximation is be to produce the correct behaviour in particular situations (e.g. the zero temperature limit) and to implement marginal stability.

In the simplest version of this approach we use the same probability distribution for $h_\alpha$ as in the SK model, that depends only the function $q(x)$. However one should use the  full Bethe form of the free energy (\ref{FE}), not the linearized version that is valid in SK limit, that depends only on the variance of the $h_\alpha$.
In this way, one obtains a relatively  simple functional $\cF^{(z)}[q]$, that in the limit $z\to\infty$ reduces the well-studied functional for the SK model. 

 Let us discuss the numerical steps  needed to implement this proposal. We  assume that the distribution of the $h_\alpha$ in the Bethe lattice is the same as the one $h_\alpha$ in the SK model for a given choice of the covariances $q(x)$. We can now proceed as follows.
 \begin{itemize}\con
\item We fix a high value of $r$, in such a way that we are {\sl near} to the limit $r=\infty$. We extract an $\epsilon$-pruned tree, e.g. using the algorithm of \cite{PRY}: the tree will have $L$ leaves, $L$ being a random quantity that diverges when $\epsilon \to 0$.
 Next, we  generate  $K$ times\footnote{The variable $K$ is as usual a random Poisson number with average $z$.}
   $h_\alpha$ ($\alpha=1,L$) using the same formula as in the SK model for the same  tree. This could be done or  using the aforementioned differential equations that are exact  in the SK model or using the stochastic differential equation.
 \item We finally compute a contribution to the free energy using eq. (\ref{FE}): the final result is  obtained by averaging the single contributions over the pruned random  tree, the variable $K$, the $LK$ fields and the random coupling. 
 \item Once that we have the final result for the free energy, we should find its maximum. The fastest way  would be to differentiate the free energy with respect  to $q(x)$ and to find the value of $q(x)$ where the derivatives are zero.
 \end{itemize}
  The implementation of this algorithm may be boring, but it is feasible.

The moderate complexity of this approach should make both numerical and analytic computations feasible. The cavity equations are likely not satisfied because the true maximum does not belong to   this {\sl Markov} subspace. It would be very interesting to quantify the violations of the cavity equations and understand if they are small. The most important check would be to verify if marginal stability is not spoiled by these approximations.

I would like to stress that this simple variational approach will produce values for the fields that cannot be correct: their distribution is continuous, while in the zero temperature limit we must have that $h$ takes integer values in the model with $J=\pm 1$. However, after one (manageable) iteration of the cavity equations, the fields should take  integer values.
\subsection{Miscellaneous remarks}

The previous equations  imply  simple expressions for the conjoint probabilities of the fields. Let us consider for example the case of $s$ states that have the same common ancestor at $x$. The conjoint probability of the fields $h_t$ (where $t$ is convenient label for these states that ranges from 1 to $s$) can be factorized as
\begin{equation}
p(h_1,h_2 \dots h_s)=\int dh P(h,x|0,0)\, P(h_1|h,x)P(h_2|h,x)\dots P(h_s|h,x)\, . \label{Multiple}
\end{equation}
This factorization property is indeed valid in the SK model\footnote{In the SK model $A(x,z)$ does not depends on $z$ and $C(z)=z$.}. 

I would like to stress that at this stage the quantity $h(x|x_0,h_0)_\eta$ is not an effective field, but a convenient label to parametrize the probability distribution in the previous equation (\ref{Multiple}). It is not clear to me if  $h(x|x_0,h_0)_\eta$ has a deeper  simple physical meaning\footnote{
If we compute the annealed free energy $F(x)\equiv \lim_{N\to\infty}-{\log\left(\overline{Z^N_J(\beta)^x}\right)/(\beta N x)}$ in presence of a magnetic field, we can derive similar formulae where the quantity $h(x|x_0,h_0)_\eta$ has a more important physical role.
}. On the contrary, $m(h,x)$ is the average magnetization over the states 
%(the averaging has to be done not including the weights $w_\alpha$) 
 that belong to the cluster that starts from $x$ and that are labeled by $h$.

 It is quite likely  that this factorized representation is not exact. One could to consider more complex representations as 
\begin{equation}
p(h_1,h_2\dots h_s)=\int d^Dz P(\vec{z}|x)P(h_1|\vec{z},x)P(h_2|\vec{z},x)\dots P(h_s|\vec{z},x) \, ,
\end{equation}
where $\vec{z}$ is a $D$ dimensional vector. It is possible that such representations are more accurate and maybe correct in the limit $D\to \infty$.

In a similar way can generalize the previous  considerations introducing a $D$ dimensional white noise and writing   the following stochastic differential equations:\begin{equation}
{d \vec{z}_{\vec{\eta}}(x) \over dx}= A\left(x,\vec{z}_{\vec{\eta}}(x)\right)\eta(x)+B(x,\vec{z}_{\vec{\eta}}(x)) \, .
\end{equation}
As in the previous case  we set $h_\alpha=C(\vec{z}_{\vec{\eta}})$. In other words we have a Markov type property in a $D$ dimensional space.

In this way, we obtain a new functional that has to maximize: increasing $D$ we should get better results. We wonder if  in the infinite $D$ limit we obtain the correct result. However, the real point is how manageable  small $D$ approximations (mainly $D=1$) do capture the crucial physics of the system. 

We have already remarked that in the one step case (i.e. $r=1$) for the model with couplings $J=\pm 1$ in the zero temperature limit the probability distribution of the fields $h$ in a given state can be described by the probabilities that $h>0$, $h=0$, $h<0$. These three probabilities (i.e surveys \cite{MP2,MPZ}) form a two-dimensional manifold. Therefore it is natural to introduce a $D=2$ parametrization in this case. It would be  interesting to check what happens in the zero temperature limit if we make the extra assumption that the probability of the surveys is concentrated on a one-dimensional manifold. It particular it would be nice to know how much the bounds on the free energy  worsen in this case.

\subsection{An analytic approach}

We can also attack this problem more analytically.
Bethe lattice spin glasses  behave as the  SK model both in the limit $z \to \infty$ or near the critical temperature: we know that in this case $D=1$ gives the correct results.

The perturbative corrections in $1/z$ or $T_c-T$ can be computed analytically. In this framework, we can investigate  how much accurate results the previous Markov ansatz does  provide.

We notice in some case that these expansions could also be used to obtain the needed results and also to check that the qualitative properties we are interested in are not different from those of the SK model. 

In the Sherrington-Kirkpatrick case, the free energy can be written in terms the covariance of the $h$'s.
 The connected correlation functions of  two more $h$'s can be computed using equations like (\ref{Multiple}), but these correlation functions do not appear in the computation of the free energy. In both cases ($z\to\infty$ and $T\to T_c$) the contributions higher order moments of the probability distribution  function can be introduced perturbatively.

These two expansions are complementary one to the other:
\begin{itemize} \con
\item The expansion in powers of $T_c-T$ is particularly simple so that very detailed computation can be done. In principle, we can do the whole computation by hand. The first terms of the expansion have been computed, at least up to $(T_c-T)^4$ \cite{PR}. Using algebraic manipulators the number of terms that can be computed strongly increases: it is unclear how difficult would be to compute enough terms in such a way to have a good control of the convergence of the expansion for small $T$.

Another very interesting computations would be to write down a Taylor expansion of $g[\theta]$ in power in powers of $\theta$:
\begin{equation}
g[\theta]=\int dx_1 g_1(x_1)\theta(x_1)+\int dx_1\int dx_2 g_2(x_1,x_2)\theta(x_1) \theta(x_2)+ \cdots
\end{equation}
 If we consider  the model where the coupling $J$ have a Gaussian distribution it is possible that this expansion holds near $T_c$ and that the various $g_n$ functions  could be computed analytically in a Taylor expansion in powers of $(T_c-T)^4$.
\item The expansion in powers of $1/z$  at fixed temperature is more complex, because the solution of the model at $z=\infty$ is complex and the function $q(x)$ is only known numerically \cite{PT,Boschi}. On the other hands, typical three-dimensional lattices have $z=6$ (sc) of $z=12$ (fcc): if we want to control the effect of having a finite number of neighbours in three dimensions (neglecting the effect of spatial correlations) the value of $1/z$ is quite small and one or two terms of the expansions may be adequate.
\end{itemize}
Many interesting analyses and computations are possible and the comparison of the results coming from different approaches could be quite fruitful. I hope  this analysis will be done in the near future.

\section*{Acknowledgements}

It is a pleasure to thank Dmitry Panchenko for many discussions and clarifications. I would like also to thank Valerio Astuti, Silvio Franz, Carlo Lucibello, Federico Ricci-Tersenghi, and Pierfrancesco Urbani for a critical reading of the manuscript.
This work has  been supported by the European Research Council (ERC) under the European Unions Horizon 2020 research and innovation program (grant agreement No [694925]).

\section*{Appendix: the cavity equations in the case of broken replica symmetry}
The cavity equations are often a direct way to find the maximum of the free energy.

At this end it is convenient to introduce the cavity equations for the probability over the descriptors i.e.  $\cP_{}[\cD]$.   These equations can be derived heuristically by considering a $N$ nodes system to which we add an extra node. A natural requirement is that the properties of the spin the extra node is the same of the others $N$ spins.  Alternatively, we could remove one spin from a $N+1$ system  forming a cavity, hence the name ``cavity" method.

It has been proved \cite{PANCAV}  that a stochastically stable  probability $\cP_{}[\cD]$ that maximizes the free energy, should also satisfy of some equations that generalize the  naive cavity equations (\ref{NCE}). They are defined as follows.
 For each state $\alpha$ we construct a new set of fields and weights using the fields $vh_\alpha$ that are uncorrelated in $i$ and correlated in $\alpha$.
\begin{equation}
h'_\alpha=U(\vJ,\vh_\alpha) \ \ \  w'_\alpha={w_\alpha\Delta ^{node}_K(\vJ,\vh_\alpha)\over Z}\, , \label{HCE}
\end{equation}
$Z$ being a normalisation factor that enforces $\sum w'_\alpha=1$. The quantities $h'_\alpha$ and $w'_\alpha$ are the fields and weights  corresponding to new spin. These equations can be formally written as
\begin{equation}
\cP'= {\cal C}[\cP] \,.
\end{equation}

 According to   the heuristic argument we finally impose that the probability distribution of the $\{w'_\alpha,h'_\alpha\}$ (i.e. $\cP'$ is again $\cP[\cD]$. This constraint is called the cavity equations for  $\cP_{}[\cD]$ \cite{MP,ASS,PanBethe}: the cavity equations ($\cP'= {\cal C}[\cP]$) are valid in general for stochastically stable distributions, not only in the reproducible case.

The cavity equations can be considered the moral equivalent of 
\begin{equation}
{\delta \cF[\cP_{}[\cD]]\over \delta \cP_{}[\cD]}=0 \, . \label{FormalCavity}
\end{equation}
These equations are  a necessary condition but not a sufficient condition. The cavity equations have many solutions; we need to find out the one that maximizes the free energy $\cF[\cP_{}[\cD]]$.

In the heuristic approach one often requires a further condition of stability. It can formally written as follows. Let us consider a solution ($\cP^*[\cD]])$ of the cavity equation  (\ref{FormalCavity}). For small $\delta \cP[\cD]$ we can write a Taylor expansion
\begin{equation}
\cF[\cP^*[\cD]+\delta \cP[\cD]]=\cF[\cP^*[\cD]]+(\delta \cP[\cD], {\bf\large H}\, \delta \cP[\cD]) +O(\delta \cP[\cD]^3) \,,
\end{equation}
where the linear term in $\delta \cP[\cD]$ is absent as  effect of the cavity equation  (\ref{FormalCavity}).

This equation defines a generalized Hessian (i.e. ${\bf\large H}$) that plays a crucial role in the heuristic approach. 

Let us call  ${\large H}_0$ the largest eigenvalues of ${\bf H}$ (the so called replicon eigenvalue). It natural to assume that if $\cP^*[\cD]$  maximize the free energy only if
\begin{equation}
{\large H}_0\le 0\,,
\end{equation}
or equivalently the matrix ${\bf H}$ is non-positive,
 This condition generalizes the De Almeida-Thouless (DAT) stability condition. It is interesting to note that the DAT stability condition is usually equivalent to the condition:
 \begin{equation}
\lambda \equiv \lim_{l\to\infty}\frac{1}{l}\log\left({\overline {(\langle \sigma_0 \sigma_l\rangle_c)^2 }\over z^l}\right)\le 0\,,
\end{equation} 
where $\langle \sigma_0 \sigma_l\rangle_c$ is the connected correlation function of two spins at distance $l$ on the infinite Bethe Lattice \cite{CORRE}.
 
  It is also conjectured that this stability condition in saturated when the replica symmetry is spontaneously broken in a continuous way: 
   \begin{equation}
 {\large H}_0=0 \ \ \ \mbox{and } \ \ \ \ \lambda=0\,. \label{MS}
\end{equation} 
Indeed in all known cases, the saturation of DAT stability condition is the necessary condition for marginal stability.

It is conjectured that for spin glasses on the Bethe lattice when replica symmetry is spontaneously broken at $r$ steps we  have 
\begin{equation}
 {\large H}_0>0 \,.
\end{equation} 
This conjecture implies that the  stability condition forces us to consider continuous RSB.   This is the reason we are interested in finding  approximations to spin glasses such that equation (\ref{MS}) is satisfied, maybe in some restricted space.
%
%
%
%Let us call  ${\large H}_0$is the largest eigenvalues of ${\bf\large H}$ (the so-called replicon eigenvalue). it is usually conjectured that if $\cP^*[\cD]$  maximize the free energy only if
%\begin{equation}
%z{\large H}_0\le 1\,.
%\end{equation}
% This condition generalizes the De Almeida-Thouless (DAT) stability condition. It is also conjectured that this stability condition in saturated in the cases where the replica symmetry is spontaneously broken in a continuous way: in this cases:
% \begin{equation}
%z {\large H}_0=1 \,. \label{MS}
%\end{equation} 
%Indeed in all know cases the saturation of DAT stability condition is the necessary condition for marginal stability.
%
%It is conjectured that for spin glasses on the Bethe lattice, when replica symmetry is spontaneously broken at $r$ steps we  usually 
%\begin{equation}
%z {\large H}_0>1 \,.
%\end{equation} 
%This conjecture implies that the DAT stability condition forces us to consider continuous RSB.   It would be fair to say that this article aims to a path for arriving at  approximations to spin glasses such that equation (\ref{MS}) is satisfied, maybe in some restricted space.
%

\end{document}